\documentclass[12pt]{article}
\usepackage{amsmath,amssymb}
\usepackage{latexsym}
\newtheorem{assumption}{Assumption}[section]
\newtheorem{theorem}[assumption]{Theorem}

\newtheorem{lemma}[assumption]{Lemma}

\begin{document}
\title{Constraints, gauge symmetries, and noncommutative gravity in
two dimensions}
\author{Dmitri V. Vassilevich\thanks{On leave from Department of Physics,
St.~Petersburg State University, Russia. 
}\\ {\it Institut f\"{u}r Theoretische Physik,
Universit\"{a}t Leipzig,}\\ {\it Augustusplatz 10, D-04109 Leipzig, Germany }
\\{email: \texttt{Dmitri.Vassilevich@itp.uni-leipzig.de}}}
\date{ }
\maketitle
\begin{abstract}
After an introduction into the subject we show how one constructs a
canonical formalism in space-time noncommutative theories which allows
to define the notion of first-class constraints and to analyse gauge
symmetries. We use this formalism to perform a noncommutative deformation
of two-dimensional string gravity (also known as Witten black hole).\\
\textit{Dedicated to Yu.~V.~Novozhilov on the occasion of his 80th birthday}
\end{abstract}
\section{Introduction}
Over the past decade considerable progress 
has been achieved in noncommutative field theories
\cite{NCrevs}.
These theories are defined on a manifold whose coordinates do not
commute. There are two essentially equivalent ways to describe noncommutative
coordinates. One either introduces operators instead of numbers, or
defines a new product of functions on the manifold. Here we shall use the
latter approach.

Noncommutativity is not a purely theoretical invention. Noncommutative
coordinates is a feature of many physical systems. As an example one
may consider electrons in an external magnetic field. If one then restricts
the electrons to several lowest Landau level, one gets second class 
constraints. Dirac brackets of the coordinates are then nonzero.
This situations is realized in the Quantum Hall Effect. Another important
example comes from string theory. It has been demonstrated 
that coordinates of the end points of open string do not commute.
Consequently, field theories on Dirichlet branes are noncommutative
field theories.
One can argue by using very general arguments \cite{Doplicher:1994zv}
that already classical
gravity implies noncommutativity of coordinates at short distances.

Let us now define the star product of functions which will replace
usual point-wise product.
Consider a space-time manifold $\mathcal{M}$ of dimension $D$.
The Moyal star product of functions on $\mathcal{M}$ reads
\begin{equation}
f\star g = f(x) \exp \left( \frac i2 \, \theta^{\mu\nu}
\overleftarrow{\partial}_\mu \overrightarrow{\partial}_\nu \right)
g(x) \,.\label{starprod}
\end{equation}
$\theta$ is a constant antisymmetric matrix.
This product is associative, $(f\star g)\star h=f\star (g\star h)$.
In this form the star product
has to be applied to plane waves and then extended
to all (square integrable) functions by means of the Fourier series.
Obviously,
\begin{equation}
x^\mu \star x^\nu - x^\nu \star x^\mu = i \theta^{\mu\nu} \,.
\label{xmxn}
\end{equation}
 We impose no restrictions on
$\theta$, i.e. we allow for the space-time noncommutativity.

The Moyal product is closed,
\begin{equation}
\int_{\mathcal{M}} d^Dx f\star  
g=\int_{\mathcal{M}} d^Dx f\times  g \label{clo}
\end{equation}
(where $\times$ denotes usual commutative product),
it respects the Leibniz rule
\begin{equation}
\partial_\mu (f\star g)=(\partial_\mu f)\star g + f\star (\partial_\mu g),
\label{Leib}
\end{equation}
and allows to make cyclic permutations under the integral
\begin{equation}
\int_{\mathcal{M}} d^Dx f\star  g \star h 
=\int_{\mathcal{M}} d^Dx h\star f\star  g\,.
\label{cyper}
\end{equation}
The product (\ref{starprod}) is not the only possible choice of an
associative noncommutative product. The right hand side of (\ref{xmxn})
can depend, in principle, on the coordinates.

To construct a noncommutative counterpart of given commutative 
field theory on has to replace all point-wise products by the star
products. The result is, of course, not unique. There are some natural
restrictions on noncommutative deformations of field theories.
For example, one usually requires that number of gauge symmetries
is preserved by the deformation.

Among all noncommutative field theories
the theories with space-time noncommutativity have a somewhat lower
standing since it is believed that they cannot be properly quantised
because of the problems with causality and unitarity (see, e.g.,
\cite{problems}). Such problems occur due to the time-nonlocality
of these theories caused by the presence of an infinite number of
temporal derivatives in the Moyal star product. However, it has been
shown later, that unitarity can be restored 
\cite{unirest}\footnote{One has to note that the approach based on
time-ordered perturbation theory has some internal difficulties
\cite{Reichenbach:2004ca}.}
(see also \cite{Balachandran:2004rq})
in space-time noncommutative theories and that the path integral
quantisation can be performed \cite{Fujikawa:2004rt}.
This progress suggests that space-time noncommutative theories
may be incorporated in general formalism of canonical quantisation
\cite{cabooks}. Indeed, a canonical approach has been
suggested in \cite{Gomis:2000gy}.

Apart from quantisation, there is another context in which canonical
approach is very useful. This is the canonical analysis of constraints
and corresponding gauge symmetries \cite{cabooks}. The problem of
symmetries becomes extremely complicated in noncommutative theories.
Already at the level of global symmetries one see phenomena which
never appear in the commutative theories. For example,
the energy-momentum tensor in translation-invariant noncommutative
theories is not locally conserved (cf. pedagogical comments in
\cite{Gerhold:2000ik}). At the same time all-order renormalizable
noncommutative $\phi^4$ theory is 
\emph{not} translation-invariant \cite{GrWu}.
A Lorentz-invariant interpretation of noncommutative space-time leads
to a twisted Poincare symmetry \cite{Chaichian:2004za}.
It is unclear how (and if) this global symmetry can be related to local
diffeomorphism transformations analysed, e.g., in \cite{Jackiw:2001jb}.
Proper deformation of gauge symmetries of generic two-dimensional
dilaton gravities remains on open problem (see below).  
Solving (some of) the problems related to gauge symmetries 
in noncommutative field theories by the canonical methods is the main 
motivation for this work.

We start our analysis from the very beginning, i.e. with a definition of the
canonical bracket. Our approach is based on two main ideas.
First of all, we separate implicit time derivatives (which are
contained in the Moyal star), and explicit ones (which survive in
the commutative limit). Only explicit derivatives define the canonical
structure. As a consequence, the constraints and the hamiltonian
become non-local in time. Therefore, the notion of same-time
canonical brackets becomes meaningless. We simply postulate a
bracket between canonical variables taken at different
points of space ($\mathbf{x}$ and $\mathbf{x}'$) and of time
($t$ and $t'$): 
\begin{equation}
\{q_a(\mathbf{x},t),p^b(\mathbf{x}',t')\} =\delta_a^b \delta (\mathbf{x}-
\mathbf{x}')\delta (t-t')\label{stbra}
\end{equation}
This bracket is somewhat similar to the one appearing in
the Ostrogradski formalism for theories with
higher order time derivatives (see, e.g., \cite{Ostro} 
for applications to field theories
and  \cite{Gomis:2000gy} for the use in space-time noncommutative theories),
but there are important differences (a more detailed comparison
is postponed until sec.\ 3).  

Of course, the proposed formalism means a departure
from the standard canonical procedure. Nevertheless, we are able to
demonstrate that the new bracket satisfies such fundamental requirements
as antisymmetry and the Jacobi identities. These brackets generate 
equations of motion. Moreover, one can define the notion of
first-class constraints with respect to the new bracket and
show that these constraints generate gauge symmetries of the action.
We shall derive an explicit form of the symmetry transformation
and see that they look very similar to the commutative case
(the only difference, in fact, is the modified bracket and the
star product everywhere). We stress that our bracket will be used
here to analyse gauge symmetries of classical systems only.
It is not clear whether such a bracket is useful for quantisation.

The main application of the canonical formalism proposed here 
is noncommutative gravity theories in two dimensions. Let us consider
the commutative case first (see review \cite{Grumiller:2002nm} where
one can also find a more extensive literature survey). 
Since the Einstein-Hilbert Lagrangian density
 in two dimensions is a total derivative, one has to introduce a scalar field
$\phi$ (called dilaton) so that the action reads: 
\begin{equation}
S=\int d^2x \sqrt{-g} \left[ \frac R2 \phi -\frac{U(\phi)}2 (\nabla\phi)^2
+V(\phi) \right]\,. \label{2ndord}
\end{equation}
This action is general enough to describe many important gravity
theories in two dimensions. For example, the choice
\begin{equation}
V(\phi)=\Lambda \phi \,,\qquad U(\phi)=0 \label{JTpot}
\end{equation}
yields the Jackiw-Teitelboim (JT) model \cite{JT}\footnote{The equations 
of motion for this model were first studied in \cite{Barbashov:1980bm}.}. 
Spherically
symmetric reduction of the Einstein theory in $D$ dimensions
leads to the dilaton gravity action in two dimensions with the potentials: 
\begin{equation}
V(\phi) \propto \phi^{\frac{D-4}{D-2}} \,,\qquad 
U(\phi) \propto \frac 1\phi \,.
\label{SRGpot}
\end{equation}
The low energy limit of string theory \cite{stringgrav} will be of particular
importance for the present work. It is described by the potentials:
\begin{equation}
V(\phi)=-2\lambda^2 \phi, \qquad U(\phi) = -\frac 1\phi\,. \label{CGHSpot}
\end{equation}
This model is also called the Witten black hole.

By a dilaton dependent conformal transformation 
$g_{\mu\nu}=e^{-2\rho}\tilde g_{\mu\nu}$ with 
\begin{equation}
\rho =-\frac 12 \int^\phi U(Y) dY \label{rhoX}
\end{equation}
one obtains an action for the metric $\tilde g$ again in the
form (\ref{2ndord}) but with the potentials
\begin{equation}
\tilde U=0\,,\qquad \tilde V=V\exp (-2\rho) \,.\label{tilUV}
\end{equation}
For the string gravity (\ref{CGHSpot}) the potential
\begin{equation}
\tilde V=-2\lambda^2 \label{tilCGHS}
\end{equation}
is a constant. Note, that the transformation $g\to \tilde g$ may be
singular, so that conformally related theories describe, in general,
globally inequivalent geometries. However, this conformal transformation
may be very useful as it simplifies the local dynamics considerably.

The action (\ref{2ndord}) can be rewritten in the first order form:
\begin{equation}
S=\int \left[ \phi_a De^a +\phi d\omega 
+\epsilon \left( \frac{\phi_a \phi^a}2 U(\phi) +V(\phi)
\right)\right]\,,
\label{1stord}
\end{equation}
where we have used the Cartan notations, $e^a=e^a_\mu dx^\mu$ is the 
zweibein one-form, $a=0,1$ is the Lorentz index, 
$\omega=\omega_\mu dx^\mu$ is the
spin-connection one-form (usual spin-connection is 
$\omega_\mu \varepsilon^{ab}$, with $\varepsilon^{ab}$ being the
Levi-Civita symbol). $\epsilon$ is the volume two-form. 
$De^a=de^a +{\varepsilon^a}_b\, \omega \wedge e^b$ is the torsion two-form. 
To prove the equivalence \cite{Katanaev:1995bh}
one has to exclude auxiliary fields $\phi_a$ and
the torsion part of $\omega$ by means of algebraic equations
of motion. The rest then depends on $e^a$ only through the
metric $g_{\mu\nu}=e_\mu^a e_{\nu a}$ and is indeed equivalent
to (\ref{2ndord}). The proof of quantum equivalence 
\cite{Kummer:1996hy} is more tricky.

Commutative dilaton gravities in two dimensions are being
successfully used to get an insight into such complicated
problems as gravitational collapse, information paradox,
and quantisation of gravity. In the noncommutative
case only the JT model was treated in some detail in classical
\cite{Cacciatori:2002ib} and quantum 
\cite{Vassilevich:2004ym} regimes. We also like to mention an
alternative approach  \cite{Buric:2004rm} to noncommutative geometry
in two dimensions which does not use any particular action.

In this paper we construct another two-dimensional noncommutative
dilaton gravity which is a deformation of conformally transformed
string gravity and analyse its gauge symmetries by using the
canonical analysis suggested below.
\section{Canonical bracket}
The phase space on $\mathcal{M}$ consists of the variables
$r_j$ which can be subdivided into canonical pairs $q,p$ and other
variables $\alpha$ which do not have canonical partners (these will
play the role of Lagrange multipliers or of gauge parameters).
We define a bracket $(r_j,r_k)$ to be $\pm 1$ on the canonical pairs,
\begin{equation}
(q_a,p^b)=-(p^b,q_a)=\delta_a^b \label{rbr}
\end{equation}
and zero otherwise (e.g., $(\alpha,p)=(p^a,p^b)=0$).
With this definition the bracket (\ref{stbra}) reads:
$\{r_i(x),r_j(x')\}=(r_i,r_j)\delta (x-x')$. Note, that we are not going
to use brackets between two local expressions (see discussion below).

Now we can define canonical brackets between star-local functionals
on the phase space. We define 
the space of star-local \emph{expressions} as
a suitable closure of the space 
of free polynomials of the phase space variables $r_j$ and their
derivatives evaluated with the Moyal star. Such expressions integrated over 
$\mathcal{M}$ we call star-local \emph{functionals}. 

Locality plays no important role here, since after the closure one can
arrive at expressions with arbitrary number of explicit derivatives
(besides the ones present implicitly through the Moyal star).
It is important, that all expressions can be approximated with
only one type of the product (namely, the Moyal one), and no
mixed expressions with star and ordinary products appear.
One also has to define what does ``suitable closure'' actually mean,
i.e. to fix a topology on the space of the functionals.
This question is related to the restrictions which one imposes on
the phase space variables. For example, the bracket of two
admissible functionals (see (\ref{canobr}) below) should be
again an admissible functional. This implies that all integrands
are well-defined and all integrals are convergent. Stronger restrictions
on the phase space variables mean weaker restrictions on the
functionals, and vice versa. Such an analysis cannot be done
without saying some words about $\mathcal{M}$ (or about its'
compactness, at least)\footnote{Some restrictions on $\mathcal{M}$
follow already from the existence of the Moyal product,
which requires existence of a global coordinate system at least
in the noncommutative directions.}. 
We shall not attempt to do this analysis here
(postponing it to a future work). All statements made below are
true at least for $r\in C^\infty$ and for polynomial functionals
(no closure at all). 

Obviously, it is enough to define the bracket
on monomial functionals and extend it to the whole space by
the linearity. Generically, two such monomial functionals read:
\begin{equation}
R=\int d^Dx\, \partial_{\kappa_1} r_1\star 
\partial_{\kappa_2} r_2 \star \dots
\partial_{\kappa_n} r_n \,,\qquad
\tilde R=\int d^Dx\, \partial_{\tilde \kappa_1}  
\tilde r_1 \star \partial_{\tilde \kappa_2} \tilde r_2 \star\dots
\partial_{\tilde \kappa_m}\tilde r_m \label{RtR}
\end{equation}
$\kappa_j$ is a multi-index, $\partial_{\kappa_j}$ is a differential
operator of order $|\kappa_j|$.
The (modified) canonical bracket of two monomials is defined by the equation
\begin{eqnarray}
&&\{ R,\tilde R \}=
\sum_{i,j} \int d^Dx\, \partial_{\kappa_j} \left( \partial_{\kappa_{j+1}}
r_{j+1} \star \dots \partial_{\kappa_{j-1}}
r_{j-1} \right) (r_j,\tilde r_i) \nonumber\\
&&\qquad\qquad\qquad \star
\partial_{\tilde\kappa_i} \left( \partial_{\tilde \kappa_{i+1}}
\tilde r_{i+1}\star \dots \partial_{\tilde \kappa_{i-1}}
\tilde r_{i-1} \right) (-1)^{|\kappa_j|+|\tilde \kappa_i|}\,.
\label{canobr}
\end{eqnarray}
In other words, to calculate the bracket between two monomials one has
to (i) take all pairs $r_j$, $\tilde r_i$; (ii) use cyclic
permutations under the integrals to move $r_j$ to the last place,
and $\tilde r_i$ -- to the first; 
(iii) integrate by parts to remove derivatives
from $r_j$ and $\tilde r_i$; (iv) delete $r_j$ and $\tilde r_i$,
put the integrands one after the other connected by $\star$
and multiplied by $(r_j,\tilde r_i)$; (v) integrate over 
$\mathcal{M}$. Actually, this is exactly the procedure one uses
in usual commutative theories modulo ordering ambiguities following
from the noncommutativity. 

The following Theorem demonstrates that the operation we have
just defined gives indeed a Poisson structure on the space of star-local
functionals.  
\begin{theorem} \label{ThPoi}
Let $R$, $\tilde R$ and $\hat R$ be star-local
functionals on the phase space. Then\\
\textrm{(1)} $\{ R,\tilde R\}=-\{ \tilde R,R\}$ (antisymmetry),\\
\textrm{(2)} $\{ \{ R ,\tilde R\}, \hat R\}+
 \{ \{\hat R , R\}, \tilde R\}+\{ \{\tilde R ,\hat R\}, R\}=0 $
(Jacobi identity).
\end{theorem}
\noindent \textbf{Proof}. We start with noting that since we do
not specify the origin of the canonical variables, the time
coordinate does not play any significant role, and the statements
above (almost) follow from the standard analysis \cite{cabooks}.
However, it is instructive to present here a complete proof as
it shows that one do not need to rewrite the star product through
infinite series of derivatives (so that the $\star$ product indeed
plays a role of multiplication). Again, it is enough to study the
case when all functionals are monomial ones. Then the first assertion
follows from (\ref{canobr}) and $(r_j,r_k)=-(r_k,r_j)$. Let
\begin{equation}
\hat R=\int d^Dx\, \partial_{\hat \kappa_1} 
\hat r_1 \star \partial_{\hat \kappa_2} \hat r_2 \star \dots
\partial_{\hat \kappa_p}\hat r_p \,. \label{hatR}
\end{equation}
Consider $ \{ \{ R , \tilde R\}, \hat R\}$. The first of the brackets 
``uses up'' an $r_j$ and an $\tilde r_i$. The second bracket uses
a variable with hat and another variable either from $R$ or from
$\tilde R$. Consider first the terms in the repeated bracket 
which use twice some variables from $R$. All such terms combine into
the sum
\begin{eqnarray}
&&\sum_{i,k,j\ne l} (-1)^{|\tilde \kappa_i|+|\hat \kappa_k|}
(r_j,\tilde r_i)(r_l,\hat r_k) \int d^Dx\, 
\partial_{\kappa_{l+1}}r_{l+1}\star \dots \partial_{\kappa_{j-1}}r_{j-1}
\nonumber\\
&&\quad\star \partial_{\kappa_j+\tilde\kappa_i} \left(
\partial_{\tilde \kappa_{i+1}} \tilde r_{i+1}\star \dots
\partial_{\tilde \kappa_{i-1}} \tilde r_{i-1}\right)\star 
\partial_{\kappa_{j+1}}r_{j+1}\star \dots \partial_{\kappa_{l-1}}r_{l-1}
\nonumber\\
&&\quad\star\partial_{\kappa_l + \hat\kappa_k} \left(
\partial_{\hat \kappa_{k+1}} \hat r_{k+1}\star \dots
\partial_{\hat \kappa_{k-1}} \hat r_{k-1}\right)
\nonumber
\end{eqnarray}
This complicated expression is symmetric with respect to interchanging 
the roles
of the variables with hats and the variables with tilde. Therefore, it is
clear that the terms having two brackets with $r$ in
$\{ \{\hat R , R\}, \tilde R\}$ have exactly the same form as above but
with a minus sign. No such terms (with two brackets with $r$) may appear
in $\{ \{\tilde R ,\hat R\}, R\}$. Therefore, this kind of terms are 
cancelled in $\{ \{ R ,\tilde R\}, \hat R\}+
 \{ \{\hat R , R\}, \tilde R\}+\{ \{\tilde R ,\hat R\}, R\}$. By repeating 
the same arguments for $\hat r$ and $\tilde r$ one proves
our second assertion. $\Box$

One can define a canonical bracket between functionals and densities
(star-local expressions) by the equation:
\begin{equation}
\{ R, h(r )(x) \} :=\frac{\delta}{\delta \beta (x)}
\left\{ R, \int d^Dy\, \beta (y)\star h(r)(y) \right\} \,.\label{gloloc}
\end{equation}
To construct brackets between two densities (i.e., to give a proper
extension of (\ref{stbra}) to nonlinear functions) one has to
define star-products with delta-functions which may be a very non-trivial
task. We shall never use brackets between densities.

To use the canonical bracket in computation of variations we need the
following technical Lemma.
\begin{lemma}\label{varlemma}
Let $p^a$ and $q_b$ depend smoothly on a parameter $\tau$.
We assume that the variables $\alpha (x)$ (these are the ones which do
not have canonical conjugates) do not depend on $\tau$. Let $h(r(\tau))$
be a star-local expression on the phase space. Then
\begin{eqnarray}
&&\partial_\tau \int d^Dx \beta \star h(r(\tau))=
\int d^D x\, \left( (\partial_\tau q_a)\star 
\left\{  \int d^Dy \beta \star h(r), p^a(x) \right\}\right.\nonumber\\
&&\qquad\qquad\qquad \left.- (\partial_\tau p^a)\star 
\left\{  \int d^Dy \beta \star h(r), q_a(x) \right\} \right)\label{l1}
\end{eqnarray}
\end{lemma} 
\noindent\textbf{Proof.} Obviously, it is enough to prove this Lemma
for $\beta=1$. Let us consider first the case when just one of the
canonical variables (say, $p^b$ for a just single value of $b$) 
depends on $\tau$, and when 
$h(r)=h_1(r)\star \partial_\kappa p^b\star h_2(r)$
where neither $h_1$ nor $h_2$ depend on $p^b$. Then
\begin{equation}
\partial_\tau \int d^Dx\, h(r)=
\int d^Dx\, h_1\star \partial_\kappa (\partial_\tau p^b)\star h_2
=(-1)^{|\kappa|} \int d^Dx\, \partial_\kappa (h_2\star h_1) \star
\partial_\tau p^b\,.
\label{l111}
\end{equation}
On the other hand, by using (\ref{canobr}), one obtains
\begin{equation}
\left\{ \int d^Dx\, h(r),\int d^Dy\, \beta(y)\star q_b (y) \right\}=
-(-1)^{|\kappa|} \int d^Dx\, \partial_\kappa (h_2\star h_1)\star \beta \,.
\label{l112}
\end{equation}
Next we use (\ref{gloloc}) to see that the statement of this Lemma is
indeed true for the simplified case considered. In general case
one has to sum up many individual contributions to both sides of (\ref{l1})
from different canonical
variables occupying various places in $h$.
Each of this contributions can be treated in the same way as above. $\Box$ 

As an application, consider a noncommutative field theory described by
the action
\begin{equation}
S=\int \left( p^a\partial_t q_a -h(p,q,\lambda ) \right) d^Dx
=\int p^a\partial_t q_ad^Dx -H \,,\label{Sh}
\end{equation}
where $h$ is a star-local expression,
it contains temporal derivatives 
only implicitly, i.e. only though the Moyal star. 
Note, that due to (\ref{clo})
the star between $p^a$ and $\partial_tq_a$ can be omitted.
If one takes into
account explicit time derivatives only, one can write
$p^a=\delta S/(\delta \partial_t q_a)$. Then, $H=S-\int p\partial_t q d^Dx$. 

The equations of motion generated from the action (\ref{Sh}) by
taking variations with respect to $q$ and $p$ can be written in
the ``canonical'' form:
\begin{equation}
\partial_t p^a + \{ H,p^a\}=0\,,\qquad \partial_tq_a +\{ H,q_a\}=0
\label{eom}
\end{equation}
This can be easily shown by taking $q(\tau)=q+\tau \delta q$ and
$p(\tau) = p+\tau \delta p$ and using Lemma \ref{varlemma}. No explicit
time derivative acts on $\lambda$. In a commutative theory $\lambda$
generates constraints.
\section{Constraints and gauge symmetries}
Let us specify the form of (\ref{Sh}):
\begin{equation}
S=\int \left( p^a\partial_t q_a -\lambda^j \star G_j (p,q) 
-h(p,q) \right) d^Dx
\label{SwithG}
\end{equation}
We shall call $G_j(p,q)$ a constraint, although due to the
presence of the Moyal star it cannot be interpreted as a condition
on a space-like surface. 
Dirac classification of the constraints can be also performed with the
modified canonical bracket. We say that the constraints $G_j(p,q)$ are
first-class if their brackets with $h(p,q)$ and between each other
are again constraints, i.e.,
\begin{eqnarray}
&& \left\{ \int d^Dx \alpha^i\star G_i, \int d^Dx \beta^j\star G_j
 \right\} =\int d^Dx C(p,q;\alpha,\beta)^k \star G_k \,,\label{strc}\\
&& \left\{ \int d^Dx \alpha^i \star G_i, \int d^Dx h(p,q) \right\}
=\int d^Dx B(p,q;\alpha)^k \star G_k \,.\label{withh}
\end{eqnarray} 
By Theorem \ref{ThPoi}(1) the structure functions are antisymmetric,
$C(p,q;\alpha,\beta)^j=-C(p,q;\beta,\alpha)^j$. Further restrictions
on $C$ and $B$ follow from the Jacobi identities (cf. 
Theorem \ref{ThPoi}(2)).

\begin{theorem}\label{Tgau}Let $G_i(p,q)$ be fist-class constraints
(so that (\ref{strc}) and (\ref{withh}) are satisfied). Then the 
transformations
\begin{eqnarray}
&&\delta p^a = \left\{ \int d^Dx \alpha^j\star G_j,\, p^a\right\}\label{tr1}\\
&&\delta q_b = \left\{ \int d^Dx \alpha^j\star G_j,\, q_b\right\}\label{tr2}\\
&&\delta \lambda^j=-\partial_t \alpha^j -C(p,q;\alpha,\lambda)^j
-B(p,q;\alpha)^j\label{tr3}   
\end{eqnarray}
with arbitrary $\alpha^j$ are gauge symmetries of the action (\ref{SwithG}).
\end{theorem}
\noindent\textbf{Proof.} To prove this Theorem we simply check 
invariance of (\ref{SwithG}) under (\ref{tr1}) - (\ref{tr3}).
Let $f(p,q)$ be an arbitrary star-local
expression depending on the canonical variables $p$ and $q$ only.
Then, by (\ref{tr1}) and (\ref{tr2}),
\begin{equation}
\delta\, f(p,q)
=\left\{ \int d^Dx \alpha^j \star G_j,\, f(p,q)\right\}\label{tr4}
\end{equation}
It is now obvious that that the transformations of $G$ and $h$
in the action (\ref{SwithG}) are compensated by the second and third
terms in $\delta\lambda$ respectively. The remaining term in the
action transforms as
\begin{eqnarray}
&&\delta \int d^Dx \, p^a\partial_t q_a=\nonumber\\
&&=
\int d^Dx \left( \left\{ \int d^Dy \alpha^j\star G_j,p^a(x)\right\} \star
\partial_tq_a
+ p^a \star\partial_t\left\{ \int d^Dy \alpha^j\star G_j,q_a(x)\right\}
\right)\nonumber\\
&&= 
\int d^Dx \left( \left\{ \int d^Dy \alpha^j \star G_j,p^a(x)\right\} \star
\partial_tq_a
-(\partial_t p^a)\star \left\{ \int d^Dy \alpha^j \star
 G_j,q_a(x)\right\}\right)
\nonumber\\
&&=\int d^Dx \alpha^j\star \partial_t G_j=
-\int d^Dx\partial_t( \alpha^j)\star G_j
\label{varvar}
\end{eqnarray}
Here we used integration by parts and Lemma \ref{varlemma}.
The last term in (\ref{varvar}) is compensated by the first (gradient)
term in the variation (\ref{tr3}). Therefore, the action (\ref{SwithG})
is indeed invariant under (\ref{tr1}) - (\ref{tr3}).$\Box$

Let us compare the technique developed here to the Ostrogradski
formalism for theories with higher time derivatives.
In this formalism \cite{Ostro,Gomis:2000gy}
new phase space variables $P(t,T)=p(t+T)$ and 
$Q(t,T)=q(t+T)$ are introduced. Then $t$ is interpreted as an evolution
parameter, while $T$ labels degrees of freedom (number of degrees of freedom
is proportional to the order of temporal derivatives). Then a delta-function
$\delta (T-T')$ appears naturally on the right hand side of the Poisson
brackets between $Q$ and $P$ calculated at the same value of $t$.
By returning (naively) to the original variables $q$ and $p$ one obtains 
(\ref{stbra}). In the approach of \cite{Gomis:2000gy} one proceeds in
a different way. The resulting dynamical system is interpreted as
a system with an infinite number of second-class constraints.
Additional first-class constraints would lead to considerable complications
in this procedure. It may happen that these two approaches are
equivalent, but this requires further studies.

\section{Noncommutative gravity in two dimensions}
In \cite{Vassilevich:2004sj} we considered an example of 
a two-dimensional topological
noncommutative gauge theory\footnote{In a space-space noncommutative 
theory similar calculations were done in \cite{Banerjee:2002qh}.} 
which was equivalent to a noncommutative version
\cite{Cacciatori:2002ib} of the Jackiw-Teitelboim gravity
\cite{JT}. It was the only noncommutative gravity in two dimensions
known that far. In this section we construct a new model and analyse its'
gauge symmetries.

Consider the action
\begin{equation}
S=\frac 14 \int d^2x\, \varepsilon^{\mu\nu} \left[ \phi_{ab} \star
 R_{\mu\nu}^{ab} -2\varepsilon_{ab}\Lambda e_\mu^a \star e_\nu^b  -
2\phi_a \star T_{\mu\nu}^a \right] \label{actW}
\end{equation}
with the curvature tensor
\begin{eqnarray}
&& R_{\mu\nu}^{ab}=\varepsilon^{ab} \left( \partial_\mu \omega_\nu
-\partial_\nu \omega_\mu +\frac i2 [\omega_\mu,b_\nu] +
\frac i2 [b_\mu,\omega_\nu] \right)\nonumber \\
&&\qquad +\eta^{ab} \left( i\partial_\mu b_\nu - i\partial_\nu b_\mu
+\frac 12 [\omega_\mu ,\omega_\nu] -\frac 12 [b_\mu,b_\nu ]\right)
\label{Rtensor}
\end{eqnarray}
and with the noncommutative torsion
\begin{eqnarray}
&&T_{\mu\nu}^a=\partial_\mu e^a_\nu -\partial_\nu e_\mu^a 
+\frac 12 {\varepsilon^a}_b \left( [ \omega_\mu,e_\nu^b ]_+
-[ \omega_\nu ,e^b_\mu ]_+ \right) \nonumber \\
&&\qquad\qquad\qquad\qquad\ +\frac i2 \left( [b_\mu,e_\nu^a] -
[b_\nu,e_\mu^a] \right)\,. \label{torsion} 
\end{eqnarray}
The fields $\phi$ and $\psi$ are combined into
\begin{equation}
\phi_{ab}:=\phi \varepsilon_{ab} -i\eta_{ab} \psi \,.\label{pab}
\end{equation}
Here $[\ ,\ ]_+$ denotes anticommutators. Both commutators and 
anticommutators are calculated with the Moyal star.
Noncommutative curvature and torsion were derived in \cite{Cacciatori:2002ib}.

We use the tensor $\eta^{ab}=\eta_{ab}={\mathrm{diag}}(+1,-1)$ to move
indices up and down. The Levi-Civita tensor is defined by 
$\varepsilon^{01}=-1$, so that the following relations hold
\begin{equation}
\varepsilon^{10}=\varepsilon_{01}=1,\qquad
{\varepsilon^0}_1={\varepsilon^1}_0=-{\varepsilon_0}^1
=-{\varepsilon_1}^0=1\,.\label{epsrel}
\end{equation}
These relations are valid for both $\varepsilon^{ab}$ and 
$\varepsilon^{\mu\nu}$. Note, that $\varepsilon^{\mu\nu}$ is
always used with both indices up. 

In the commutative limit the fields $b_\mu$ and $\psi$ decouple,
and the action becomes equivalent to (\ref{1stord}) with $U=0$
and $V=\tilde V$ given in (\ref{tilCGHS}).

An additional $U(1)$ gauge field is typically necessary to close
the gauge algebra in NC case. This field may play also another
role: by adding an additional abelian gauge field one can 
overcome the non-existence theorem of \cite{Grumiller:2002md} for a dilaton
action for the so-called exact string black hole \cite{Dijkgraaf:1991ba}
and construct a suitable action with the extended set of
fields \cite{Grumiller:2005sq}.

It is crucial to prove that the model (\ref{actW}) indeed has right number
of gauge symmetries. According to our analysis it is
enough to show that the constraint algebra closes w.r.t. to the
bracket defined above. One can rewrite (\ref{actW}) in 
the canonical form:
\begin{equation}
S=\int d^2x \left( p^i \partial_0 q_i - \lambda^i \star G_i \right)
\label{canact}
\end{equation}
(cf. (\ref{SwithG})).
Here:
\begin{eqnarray}
&&q_i=(e_1^a,\omega_1,b_1), \nonumber\\
&&p^i=(\phi_a,\phi,-\psi),\label{qpl}\\
&&\lambda^i=(e_0^a,\omega_0,b_0).\nonumber
\end{eqnarray}
The constraints are 
\begin{eqnarray}
&&G_a=-\partial_1\phi_a +\frac 12 {\varepsilon^b}_a[\omega_1,\phi_b]_+
+\frac i2 [\phi_a,b_1] -\Lambda \varepsilon_{ab}e_1^b \,,\label{Ga}\\
&&G_3=-\partial_1\phi +\frac i2 [\phi,b_1] +\frac i2 [\psi,\omega_1]
-\frac 12 {\varepsilon^a}_b[\phi_a,e_1^b]_+ \,,\label{G3}\\
&&G_4=\partial_1\psi -\frac i2 [\psi,b_1] +\frac i2 [\phi,\omega_1]
+\frac i2 [\phi_a,e_1^a] \,.\label{G4}
\end{eqnarray}

The following formulae hold for arbitrary trace operation
on an operator algebra. In our case, this trace is
just a space-time integral.
\begin{eqnarray}
&&\mathrm{Tr}([A_1,B_1][B_2,A_2]-[B_1,A_2][A_1,B_2])
=-\mathrm{Tr}([A_1,A_2][B_1,B_2])\label{com1}\\
&&\mathrm{Tr}([A_1,B_1]_+[A_2,B_2]_+ -[A_1,B_2]_+[A_2,B_1]_+)=
-\mathrm{Tr}([A_1,A_2][B_1,B_2])\label{com2}\\
&&\mathrm{Tr}([A_1,B_1]_+[B_2,A_2]-[B_1,A_2]_+[A_1,B_2])=
\mathrm{Tr}([B_1,B_2][A_1,A_2]_+)\label{com3}
\end{eqnarray}
These formulae help to transform the brackets into a factorized form
$\int C(\alpha,\beta)\star G(p,q)$. 
The constraint algebra indeed closes and reads
\begin{eqnarray}
&&\left\{ \int \alpha^a \star G_a,\int \beta^b \star G_b \right\}
= 0 \\
&&\left\{ \int \alpha \star G_3,\int \beta \star G_3 \right\}
= \frac i2 \int [\alpha,\beta ]\star G_4 \\
&&\left\{ \int \alpha \star G_4,\int \beta \star G_4 \right\}
=  -\frac i2 \int [\alpha,\beta ]\star G_4 \\
&&\left\{ \int \alpha \star G_3,\int \beta \star G_4 \right\}
= -\frac i2 \int [\alpha,\beta ]\star G_3 \\
&&\left\{ \int \alpha \star G_3,\int \beta^a \star G_a \right\}
= -\frac 12 \int [\alpha ,\beta^a ]_+\, {\varepsilon^b}_a \star G_b \\
&&\left\{ \int \alpha \star G_4,\int \beta^a \star G_a \right\}
= -\frac i2 \int [\alpha,\beta^a]\star G_a 
\end{eqnarray}
Here $\int :=\int d^2x$.

One can easily find gauge symmetries of the action. 
The transformations generated by $G_a$ read:
\begin{eqnarray}
&&\delta e_\mu^a=-\partial_\mu \alpha^a -\frac 12 {\varepsilon^a}_c
[\omega_\mu,\alpha_c]_+ -\frac i2 [b_1,\alpha^a],\label{Gatr}\\
&&\delta \omega_\mu =\delta b_\mu =0\,,\nonumber\\
&&\delta \phi =\frac 12 {\varepsilon^b}_a [\alpha^a,\phi_b]_+\,,
\qquad
\delta \psi =-\frac i2 [\alpha^a,\phi_a]\,,\nonumber \\
&&\delta \phi_a=-\Lambda \alpha^b \varepsilon_{ba}\,.\nonumber
\end{eqnarray}
The constraint $G_3$ generates
\begin{eqnarray}
&&\delta e_\mu^a=\frac 12 {\varepsilon^a}_b [e_\mu^b,\beta ]_+\,,
\qquad \delta b_\mu = \frac i2 [\omega_\mu, \beta ]\,,\label{G3tr}\\
&&\delta \omega_\mu = -\partial_\mu \beta -\frac i2 [b_\mu,\beta]\,,
\nonumber\\
&&\delta \phi_a=-\frac 12 {\varepsilon^c}_a[\beta,\phi_c]_+,\qquad
\delta \phi =\frac 12 [\beta,\psi ]\,,\qquad 
\delta \psi = -\frac i2 [\beta,\phi]\,.\nonumber
\end{eqnarray}
Gauge symmetries generated by $G_4$ are:
\begin{eqnarray}
&&\delta e_\mu^a=-\frac i2 [e_\mu^a,\gamma]\,,\qquad
\delta \omega_\mu =-\frac i2 [\omega_\mu,\gamma]\,,\label{G4tr}\\
&&\delta b_\mu=-\partial_\mu \gamma -\frac i2 [b_\mu,\gamma]\,,\nonumber\\
&&\delta \phi_a =\frac i2 [\gamma,\phi_a]\,,\qquad
\delta \phi =\frac i2 [\gamma,\phi]\,,\qquad
\delta \psi =\frac i2 [\gamma,\psi]\,.\nonumber
\end{eqnarray}
In (\ref{Gatr}) - (\ref{G4tr}) the functions $\alpha^a$, $\beta$
and $\gamma$ denote parameters of the gauge transformations.

In the commutative limit the transformations (\ref{Gatr}) and (\ref{G3tr})
become equivalent to diffeomorphisms and local Lorentz transformations
up to a field-dependent redefinition of the parameters
(the symmetry (\ref{G4tr}) decouples completely). Therefore, we may
say that gauge symmetries of the noncommutative action (\ref{actW}) 
contain noncommutative deformations of Lorentz and diffeomorphism
group. This is a rather nontrivial fact since $\theta_{\mu\nu}$
remains constant under the transformations. 
A more elaborate discussion on noncommutative
diffeomorphism in two dimension can be found in \cite{Cacciatori:2002ib}.
Unfortunately, it is not clear so far how one may construct
a gauge invariant line element.

To deform the Witten black hole one may use its formulation as a
Wess-Zumino-Novikov-Witten (WZWN) theory. A noncommutative formulation
of the $U(2)/U(1)$ WZWN model was constructed in \cite{Ghezelbash:2000pz}.
The paper \cite{Ghezelbash:2000pz} does not analyse gravity aspects
of the model. It remains unclear whether the deformation of
\cite{Ghezelbash:2000pz} is equivalent to the one presented above.
The action (\ref{actW}) may also be obtained as a singular limit
of the noncommutative JT model \cite{Cacciatori:2002ib}. To prove
that the gauge symmetries are preserved in this limit is of the
same level of complexity as the direct analysis presented above.

Somewhat surprisingly, construction of a proper noncommutative deformation
of classical action having proper number of gauge symmetries 
is the hardest part of the job. Analysing classical solution 
seems to be rather straightforward. Indeed, let us impose the
gauge condition
\begin{equation}
e_0^+=0,\quad e_0^-=1,\quad \omega_0=0, \quad b_0=0, \label{tempgau}
\end{equation}
where $e_\mu^\pm =2^{-1/2} \left( e_\mu^0 \pm e_\mu^1 \right)$.
Then, as one can easily see, the equations of motion become linear
and the model can be solved in a rather straightforward way.
Therefore, classical analysis of the noncommutative model considered
here is similar to what we have in the commutative case (see
\cite{Grumiller:2002nm} for more details). However, transition 
between different formulations of the dilaton gravities remains
a problem. For example, it is not clear how one should generalise
the dilaton-dependent conformal transformation described in sec.\ 1
to the noncommutative case.

The gauge condition (\ref{tempgau}) is the main technical ingredient
of exact path integral quantisation of two-dimensional commutative
dilaton gravities \cite{Kummer:1996hy}. In the case of noncommutative
JT model this gauge condition also allowed to calculate the path integral
exactly \cite{Vassilevich:2004ym}. Adding the matter fields to this
formalism \cite{Kummer:1998zs} may cause a problem.

Let us conclude this section with some remarks on possibility of
fixed background perturbative calculations in noncommutative gravity
theories. At least at one-loop order the heat kernel technique
\cite{Vassilevich:2003xt} seems to be an adequate tool.
A generalisation of the heat kernel expansion on flat Moyal
spaces was constructed recently \cite{nchk}. Even on curved Moyal
manifolds one can calculate leading heat kernel coefficients
and construct a generalisation of the Polyakov action 
\cite{Vassilevich:2004ym}. It is crucial that the operator 
describing quantum fluctuations contains only left or only right
star multiplications. If both types of multiplications are present
simultaneously, the heat kernel expansion seems to be modified in
an essential way \cite{Gayral:2004cu} due to the mixing of
ultra-violet and infra-red scales discovered previously in Feynman
diagrams \cite{UVIR}.
\section{Conclusions}\label{scon}
In this paper we have suggested a modification of the Poisson
bracket which is defined on fields at different values of the
time coordinate. In this modified canonical formalism, only
explicit time derivatives (i.e., the ones which are not hidden
in the Moyal multiplication) define the canonical structure.
Although this means serious deviations from standard canonical
methods, the resulting brackets still satisfy the Jacobi identities
(Theorem \ref{ThPoi}) and generate classical equations of motion.
Our main result (Theorem \ref{Tgau}) is that we can still define
the notion of fist-class constraints, which generate gauge symmetries,
and these symmetries are written down explicitly\footnote{Just existence
of the symmetries does not come as a great surprise in the view
of the analysis of \cite{GT} which is valid for theories with
arbitrary (but finite!) order of time derivatives. 
An important feature of the present approach is rather simple
explicit formulae similar to that in the case of commutative theories
with 1st order time derivatives.}. It would be interesting
to construct a classical BRST formalism starting with our brackets.
Anyway, it is important to restore the reputation of space-time
noncommutative theories. This is required by the principles of
symmetry between space and time, but also by interesting 
physical phenomena which appear due to the space time 
noncommutativity (just as an example we may mention creation of
bound states with hadron-like spectra \cite{Vassilevich:2003he}).
To avoid confusions we stress that our analysis is purely
classical. It is not clear whether our brackets can be used
for quantisation at all.

As an application of the canonical formalism we considered
noncommutative gravity and constructed a new deformed
dilaton gravity in two dimensions (which is a conformally
transformed string gravity). Naively one would expect that
the presence of constant $\theta_{\mu\nu}$ destrois
a part of the symmetries (and this really happens in non-gravitational
noncommutative theories). In our case, however, we observe
just right number of gauge symmetries in the deformed theory.
It seems that noncommutativity naturally leads to gravity, 
as well as gravity naturally leads to noncommutativity 
\cite{Doplicher:1994zv}.
\section*{Acknowledgements}
This paper is based on lectures given at the Advanced Summer School on
Modern Mathematical Physics (Dubna) and the International Fock School
(St.Petersburg). I am grateful to A.~T.~Filippov, V.~Yu.~Novozhilov
and Yu.~V.~Novozhilov for their kind hospitality.
I benefited from numerous discussions with Daniel Grumiller.
This work was supported in part by the DFG Project BO 1112/12-2 and by
the Multilateral Research Project "Quantum gravity, cosmology and
categorification" of the Austrian Academy of Sciences and the
National Academy of Sciences of the Ukraine.

\end{document}